\documentclass[prb,twocolumn,showpacs,preprintnumbers,amsmath,amssymb]{revtex4}

\usepackage{graphicx}
\usepackage{dcolumn}
\usepackage{bm}
\usepackage{eucal}

\begin{document}


\title{Mean-field model of the ferromagnetic ordering \\ in the
superconducting phase of ErNi$_2$B$_2$C}

\author{Jens Jensen}
 \affiliation{\O rsted Laboratory,
         Niels Bohr Institute fAPG,
         Universitetsparken 5, DK-2100 Copenhagen, Denmark}

\date{\today}

\begin{abstract}
A mean-field model explaining most of the details in the magnetic
phase diagram of ErNi$_2$B$_2$C is presented. The low-temperature
magnetic properties are found to be dominated by the appearance of
long-period commensurate structures. The stable structure at low
temperatures and zero field is found to have a period of 40 layers
along the $a$ direction, and upon cooling it undergoes a
first-order transition at $T_C^{}\simeq2.3$ K to a different
40-layered structure having a net ferromagnetic component of about
0.4 $\mu_B^{}/\hbox{Er}$. The neutron-diffraction patterns
predicted by the two 40-layered structures, above and below
$T_C^{}$, are in agreement with the observations of Choi {\it et
al.}
\end{abstract}

\pacs{74.70.Dd,75.25.+z,75.10.Jm}
\maketitle

A number of the (RE)Ni$_2$B$_2$C compounds are normal $s$-type
superconductors with a $T_c^{}$ of the order of 10 K. The
superconductors are all of type II with a $\kappa$ in the range of
6--12. They are magnetic due to the rare-earth ions and in four of
the compounds (RE = Dy, Ho, Er or Tm) the rare-earth ions are
antiferromagnetically ordered in the superconducting
phase.\cite{lynn} The ordering wave vector $\mathbf Q$ in Er
borocarbide is along an $a$ axis and has a length of about 0.55
(in units of $2\pi/a)$, and the ordered moments are along the $a$
axis perpendicular to the ordering wave vector; the N\'eel
temperature is $T_N^{}=6$ K and $T_c^{}=11$ K. The rare-earth ions
are placed in a body-centered tetragonal lattice, and in the case
of Er $a=b=3.502$ \AA\ and $c=10.558$ \AA. It was proposed already
in 1996 that the Er ions in ErNi$_2$B$_2$C develop a small
ferromagnetic component in addition to the antiferromagnetic one
below 2.3 K.\cite{canfield} This makes the Er compound
particularly interesting as presenting the case of a weak
ferromagnetic state existing well below $H_{c2}^{}$. The
ferromagnetic moment at 2 K was estimated to be 0.33 $\mu_B^{}$
per Er ion, which magnetization creates an internal magnetic field
$4\pi M\simeq 0.60$ kOe close to the estimated value of the lower
critical field $H_{c1}^{}$. This opens up the possibility for the
occurrence of exotic phenomena like a spontaneous vortex phase.
Kawano {\it et al.}\cite{kawano} did detect a ferromagnetic moment
below 2.3 K in a neutron diffraction experiment, but saw no sign
of a spontaneous vortex phase. Recently, Choi {\it et
al.}\cite{choi} have made a detail neutron-diffraction experiment
in which they measured all the higher harmonics of the
antiferromagnetic structures occurring just above and below the
Curie temperature, and they concluded that the structure at 1.3 K
has a ferromagnetic component of about 0.57 $\mu_B^{}$ per Er ion.

It is known that the superzone energy gaps on the Fermi surface
induced by the antiferromagnetic ordering may have a strong effect
on the superconducting order parameter.\cite{amici,norgaard} On
the other hand, the superconducting order parameter does not seem
to affect the antiferromagnetic ordering, but is of importance for
a ferromagnetic system at low fields. The Anderson-Suhl
mechanism,\cite{anderson} that the Ruderman-Kittel-Kasuya-Yoshida
(RKKY) of the rare-earth ions is strongly reduced in the long
wavelength limit, has been demonstrated\cite{norgaard} to be
important in TmNi$_2$B$_2$C. Here, I shall mostly concentrate on
understanding the behavior of the magnetic moments in
ErNi$_2$B$_2$C, and to start with I neglect the influence of the
superconducting ordering on the magnetic properties.

The crystal-field parameters of the Er ions have been determined
by Gasser {\it et al.} from the crystal-field transitions observed
by neutron scattering and from the high-temperature susceptibility
data.\cite{gasser} These parameters are used in the present work
except that $B_2^0$ has been scaled by a small factor (1.08), see
Table I. The ground state is a doublet, and an excited doublet is
lying only about 0.6--0.7 meV above the ground state. This
configuration leads to a four-clock behavior of the moments at low
temperatures, i.e.\ the Er ions are easily magnetized along
$\langle100\rangle$, they are hard to magnetize along the $c$
direction, and when the field is applied along $\langle110\rangle$
the moments are (approximately) a factor $\sqrt{2}$ smaller than
the moments in the $\langle100\rangle$ case.

\begin{table}[b]
 \caption{The Stevens operator parameters (meV).}
\label{table1}
\begin{ruledtabular}
\begin{tabular}{ccccc}
$B_2^0$ & $B_4^0$ & $B_4^4$ & $B_6^0$ & $B_6^4$ \\
\noalign{\vspace{1pt}} \colrule \noalign{\vspace{3pt}} $-0.0173$ \
& $0.147\cdot10^{-3}$ & $-3.3\cdot10^{-3}$ &
$-0.122\cdot10^{-5}$ & $2.16\cdot10^{-5}$\\
\end{tabular}
\end{ruledtabular}
\end{table}

Detlefs {\it et al.} have observed that the antiferromagnetic
ordering in ErNi$_2$B$_2$C is accompanied by an orthorhombic
distortion of the lattice, so that
$a/b-1=\epsilon_{11}^{}-\epsilon_{22}^{}\approx 2\cdot10^{-3}$ at
3.7 K (when $\mathbf Q$ is along the $a$ direction).\cite{detlefs}
Because of this observation, the following quadrupole coupling is
included in the model:
\begin{equation}
{\cal
H}_{\text{me}}^{}=-\sum_i\sum_{m=\pm2}K_\gamma^{m}\left[O_2^m(i)\langle
O_2^m\rangle-{\textstyle\frac{1}{2}} \langle O_2^m\rangle^2\right]
\label{1}
\end{equation}

\noindent The definition of the Stevens operators may be found in
Ref.\ [\onlinecite{jens}]. $\langle O_2^m\rangle$ is the
expectation value of the operator averaged over all ions. The
contribution to the free energy of the modulated quadrupolar
moments (at the wave vector $2{\mathbf Q}$) is a factor of 100
smaller than that of the uniform term and is neglected. The value
of $K_\gamma^{-2}$ is undetermined, but is only of minute
importance for the model and is assumed to be equal $K_\gamma^2$.
The calculated value of $\langle O_2^2\rangle$ is 34.6 at 3.7 K.
Assuming $(c_{11}^{}-c_{12}^{})/2$ to be about $5\cdot10^{11}$
ergs/cm$^3$ (consistent with a Debye temperature of the order of
400 K) then the distortion observed by Detlefs {\it et al.}\
indicates a value of $K_\gamma^2=(V/2N)(c_{11}^{}-c_{12}^{})
[(\epsilon_{11}^{}-\epsilon_{22}^{})/\langle
O_2^2\rangle]^2\approx0.7\cdot10^{-4}$ meV. This value is close to
the one used in the final fit: $K_\gamma^2=K_\gamma^{-2}=
0.8\cdot10^{-4}$ meV.

The two-ion interaction is assumed to be the sum of a Heisenberg
interaction and the classical dipole--dipole interaction:
\begin{equation}
{\cal H}_{\text{JJ}}^{}=-{\textstyle\frac{1}{2}}\sum_{i,j}
{\mathcal J}(ij){\mathbf J}_i^{}\cdot{\mathbf J}_j^{}-
{\textstyle\frac{1}{2}}\sum_{i,j} {\mathcal J}_D^{}
\,D_{\alpha\beta}^{}(ij)J_{i\alpha}^{}J_{j\beta}^{}\label{2}
\end{equation}
where the classical coupling is determined by ${\mathcal J}_D^{}=
N(g\mu_B^{})^2= 1.194$ $\mu$eV and the sum over the lattice of
\begin{equation}
D_{\alpha\beta}^{}(ij)=\frac{3(r_{i\alpha}^{}-r_{j\alpha}^{})
(r_{i\beta}^{}-r_{j\beta}^{})-|{\bf r}_i^{}-{\bf
r}_j^{}|^2\delta_{\alpha\beta}^{}}{N\,|{\bf r}_i^{}-{\bf
r}_j^{}|^5}\label{3}
\end{equation}
The two-ion Hamiltonian is accounted for in the mean-field
approximation, ${\mathbf J}_i^{}\cdot{\mathbf J}_j^{}\simeq
{\mathbf J}_i^{}\cdot\langle{\mathbf J}_j^{}\rangle
+\langle{\mathbf J}_i^{}\rangle\cdot{\mathbf J}_j^{}
-\langle{\mathbf J}_i^{}\rangle\cdot\langle{\mathbf
J}_j^{}\rangle$. All the ordered structures are described by a
wave vector $\mathbf Q$ along the $a$-axis and consist of
ferromagnetic sheets perpendicular to $\mathbf Q$. This means,
firstly, that the different positions of the ions in the two
sublattices have no direct consequences, corresponding to the use
of a double-zone representation along $\langle100\rangle$, and,
secondly, that only the total couplings between the different
ferromagnetic layers are important in the model. The interplanar
coupling parameters are
\begin{equation}
{\mathcal J}_{\parallel,\perp}^{}(n)=\sum_{{\mathbf r}_j\cdot
{\mathbf a}=na^2/2}\left[{\mathcal J}(0j)+{\mathcal J}_D^{}
D_{\parallel,\perp}^{}(0j)\right]\label{4}
\end{equation}
and the corresponding Fourier transforms ${\mathcal
J}_{\parallel,\perp}^{}({\mathbf q})$. The parameter ${\mathcal
J}_{\perp}^{}({\mathbf q})$ denotes the coupling between the
components of the moments which are lying in the $a$--$b$ plane
perpendicular to $\mathbf q$. The case of ${\mathcal
J}_{\perp}^{}(n)$ involves the following coupling parameters,
$$
{\mathcal J}_{\perp}^{}(n)={\mathcal J}_{\perp}^{}({\rm LD}
)\cos(0.558\,n \pi) \qquad;~n=10,11,\cdots,16
$$
all defined in terms of ${\mathcal J}_{\perp}^{}({\rm LD})$.
Notice, that the coupling parameter for $n=9$ is included
separately.

The final values of the interplanar coupling parameters are shown
in Table II, where ${\mathcal J}_{\perp}^{}(n)$ have been derived
from the fitting of the experimental results discussed below, and
${\mathcal J}_\parallel^{}(n)-{\mathcal J}_\perp^{}(n)$ have been
calculated using Eq.\ (\ref{3}). The results for ${\mathcal
J}_{\perp}^{}({\mathbf q})$ and ${\mathcal
J}_{\parallel}^{}({\mathbf q})$ are shown in Fig.\ 1. The most
important parameters are the wave vector at which ${\mathcal
J}_\perp^{}({\mathbf q})$ has its maximum, ${\mathbf q}={\mathbf
Q}_0^{}=0.558~{\mathbf a}^\star$, the maximum value itself
${\mathcal J}_\perp^{}({\mathbf Q}_0^{})$ and ${\mathcal
J}_\perp^{}({\mathbf 0})={\mathcal J}_\parallel^{}({\mathbf
0})=-12~\mu$eV. The value of the parameter ${\mathcal
J}_\perp^{}({\mathbf Q}_0^{})$ is assumed fixed at $21.56~\mu$eV,
which leads to a mean-field value of the N\'eel temperature
$T_N^{}=6.0$ K. The classical interaction gives rise to a large
difference ${\mathcal J}_\parallel^{}({\mathbf q})-{\mathcal
J}_\perp^{}({\mathbf q})$ of the same order of magnitude as the
Heisenberg term itself. Due to the singular behavior of the dipole
sum this difference is cancelled at zero wave vector, i.e.\
${\mathcal J}_\parallel^{}({\mathbf q})$ makes a jump by
$4\pi{\mathcal J}_D^{}=15.0$ $\mu$eV in the limit of ${\bf
q}\to{\bf0}$.

\begin{figure}[t]
\includegraphics[width=0.85\linewidth]{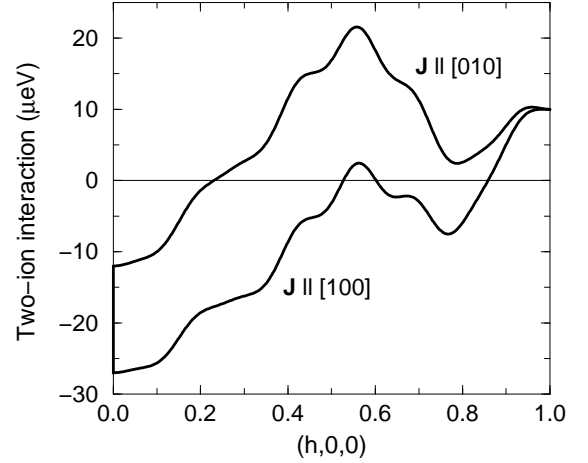}
\caption{The perpendicular and parallel components of the two-ion
interaction in ErNi$_2$B$_2$C along [100].}
\end{figure}

\begin{table*}[floatfix] \caption{The planar two-ion
coupling parameters ($\mu$eV).} \label{table2}
\begin{ruledtabular}
\begin{tabular}{crrrrrrccc}
$n$ &  0\phantom{00} & 1\phantom{00} & 2\phantom{00} &
3\phantom{00}
& 4\phantom{00} & 5\phantom{00}& 6& 9 & LD \\
${\mathcal J}_\perp^{}(n)$ & $5.847$ & $-3.816$ & $-4.786$ &
$-0.650$ & 1.500 & $-1.500$&---& 0.29 & 0.29\\
${\mathcal J}_\parallel^{}(n)-{\mathcal J}_\perp^{}(n)$ &
$-14.286$ & $-3.024$ & $3.106$ & $-0.630$ & 0.250 & $-0.084$ &
0.030
&---&---\\
\end{tabular}
\end{ruledtabular}
\end{table*}

The neutron-diffraction experiment of Choi {\it et al.}\cite{choi}
revealed a large third harmonic below $T_N^{}/2$ showing that the
modulation of the ordered moments approaches a square wave. The
system shows a number of features indicating that commensurable
effects are of decisive importance. The commensurable structures
appearing may be rather complex limiting the possibility of
deciphering the neutron-diffraction experiments. It has previously
turned out to be of great value to assist the analysis of
diffraction experiments by theoretical model calculations of the
stability of the structures which may occur. This has been the
case in the study of the long-period commensurable structures in
the elemental erbium and holmium metals.\cite{jens,cowley,jens2}
The method used here is the same as in the previous works just
cited, i.e.\ the free energies of different commensurable
structures are calculated within the mean-field approximation by a
straightforward iteration procedure, and the results are compared
to each other in order to identify the most stable structure.

The mean-field model presented above accounts for a great part of
the observed properties of ErNi$_2$B$_2$C. Figure 2 shows the
magnetization curves calculated at 2 K in comparison with the
experimental results. All the calculated results are based on
commensurable structures derived from the basic structure at
$Q=\frac{1}{2}$ (in units of $a^\star$). At low temperatures all
the moments have a magnitude of about 7.9 $\mu_B^{}$ and in one
ferromagnetic layer they are either pointing parallel (u) or
antiparallel (d) to the $b$ axis (assuming ${\mathbf
Q}\parallel{\mathbf a}$). In the $Q=\frac{1}{2}$ structure the
ferromagnetic layers perpendicular to the $a$ axis are polarized
subsequently uudduudd$\dots$ along the $a$ direction. Structures
with larger values of $Q$ are derived from this structure by a
periodic replacement of one or more of the uu (dd) double layers
with a single u (d) layer, the so-called spin-slip
structures.\cite{gibbs} Important structures in the present case
are the $Q=\frac{6}{11}$ structure consisting of the eleven
layered period d(uudd)$^2$uu = d(5p), the eighteen layered
$Q=\frac{5}{9}$ structure with the period d(uudd)$^2$u(dduu)$^2$ =
d(4p)u(4p), and the seven layered $Q=\frac{4}{7}$ structure with
the period duudduu = d(3p). Experimentally, the length of the
ordering wave vector lies in the interval between 0.554 and
0.548,\cite{choi} which range is covered by appropriate
combinations of the three genetic structures. If the moments are
of constant length, the d(5p) and the d(3p) structures have a net
ferromagnetic component equal to $\frac{1}{11}\mu_{\text{max}^{}}$
and $\frac{1}{7}\mu_{\text{max}^{}}$, respectively, whereas the
structure d(4p)u(4p) has no uniform component.

\begin{figure}[t]
\includegraphics[width=\linewidth]{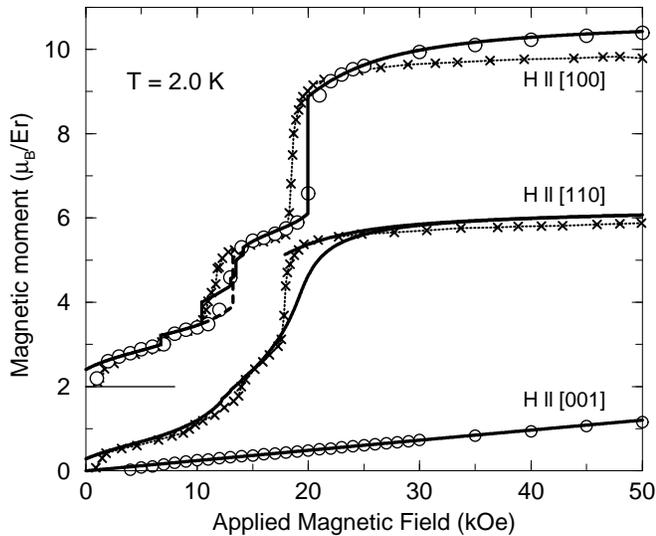}
\caption{The magnetization curves of ErNi$_2$B$_2$C at 2 K. The
open circles are the experimental results of Cho {\it et
al.}\cite{cho}. The crosses connected by dashed lines show the
experimental results of Canfield {\it et al.}\cite{canfield}. The
remaining solid and dashed lines are the calculated results. The
experimental and theoretical results in the case of
$H\parallel[100]$ have been shifted upwards by 2 units.}
\end{figure}

If the fundamental harmonic is the only one present the free
energy has its minimum value at the wave vector at which the
two-ion coupling has its maximum, i.e.\ at $Q_0=0.558$ in the
mean-field model defined above. However, the intensities of the
higher-order odd harmonics increase rapidly as the temperature is
lowered, which produce a shift of the ordering wave vector to a
smaller value. In the model $Q=0.558$ at $T_N^{}$, and it
decreases rapidly between 5 and 3 K to become about 0.55 below 3
K. This behavior is consistent with the experimental
observation\cite{choi} of a change of the ordering wave vector
between 5 and 3 K from 0.554 to 0.548. In the neutron-diffraction
experiment of Choi {\it et al.}\cite{choi} the peak due to the
fundamental harmonic is found to be centered at $Q=0.548$ at 2.4
K, however the higher harmonics indicate that the main part of the
crystal is actually ordered in a commensurable structure with
$Q=\frac{11}{20}=0.55$. At 2.4 K, the stable structure with this
wave vector is the 40-layered structure d(4p)u(5p)u(4p)d(5p). This
structure, like the d(4p)u(4p) structure above, has no
ferromagnetic component. However, due to the increasing importance
of the higher harmonics as the temperature is lowered these
structures are found to become unstable. The present mean-field
model predicts the occurrence of a first-order transition between
the structures
\begin{equation}
\hbox{d(4p)u(5p)u(4p)d(5p)}\to\hbox{d(3p)d(5p)d(5p)d(5p)}
\label{5}
\end{equation}
at 2.24 K. Numbering the layers from left to right by 1 to 40,
then the transition is accomplished by a reversal of the moments
in the layers 9 and 20. The length of the moments changes slightly
from one layer to the next, and the ferromagnetic moment is
calculated to be 0.33 $\mu_B^{}$/Er just below the transition, and
to be 0.40, 0.56 and 0.62 $\mu_B^{}$/Er at 2, 1.3 and 0 K,
respectively, in agreement with the experimental
values\cite{canfield,choi} of 0.33 $\mu_B^{}$/Er at 2 K and 0.57
$\mu_B^{}$/Er at 1.3 K. The diffraction patterns of the different
structures have been calculated and are compared with the results
of the neutron-diffraction experiment of Choi {\it et
al.}\cite{choi} in Fig.\ 3.  At 1.3 K the d(3p)d(5p)d(5p)d(5p)
configuration gives rise to both odd and even harmonics (the even
ones are marked by arrows) twice as many as produced by the
d(4p)u(5p)u(4p)d(5p) structure at 2.4 K. The only discrepancy of
some importance is the large value calculated for the intensity at
$h=0.7$ at 1.3 K. The splitting of the peaks around $h=0.45$ and
$h=0.65$ at 1.3 K may be explained if a minor part of the crystal
is ordered in the $Q=\frac{28}{51}$ structure,
d(3p)d(5p)d(5p)d(5p)d(5p). Notice, that even this small change of
the fundamental $Q$ (from 0.55 to 0.549) leads to easily
observable modifications in the positions of the higher harmonics.
Hence, the overall agreement between the observed and calculated
positions of the higher harmonics indicate with a high degree of
credibility that the main part of the crystal is a
$Q=\frac{11}{20}$ structure at 1.3 and 2.4 K. There exist other
choices for the low-temperature structure with $Q=\frac{11}{20}$
than the one proposed in (\ref{5}), but this structure, which is
calculated to be the most stable one, is the only one producing a
diffraction pattern that is reasonably similar to the one observed
at 1.3 K.

\begin{figure}[t]
\includegraphics[width=\linewidth]{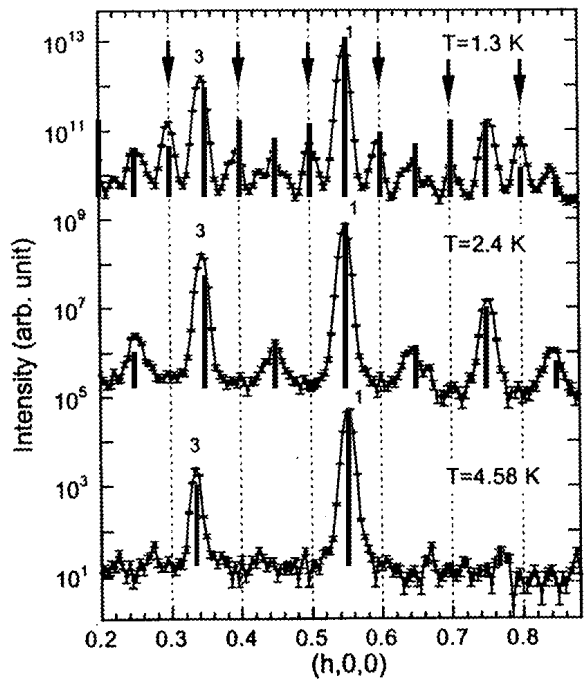}
\caption{Scan along $[h,0,0]$ at $T=1.3$ K, 2.4 K, and 4.38 K,
measured with unpolarized neutrons. The data have been offset for
clarity. The figure is a copy of Fig.\ 1 from Ref.\
[\onlinecite{choi}] in which the calculated diffraction patterns
 in the different cases (the heavy solid lines) have been incorporated. }
\end{figure}

The model calculations indicate that ${\mathcal J}({\bf q})$ has a
sharp peak at ${\bf q}={\bf Q}_0^{}$, as constructed in terms of
the parameter ${\mathcal J}({\rm LD})$. If this parameter is
neglected the calculated variation of $Q$ with temperature, or as
a function of field, increases drastically. The dependence of $Q$
on an applied field is going to be discussed in a forthcoming
paper by Toft {\it et al.}\cite{toft} In the present model all the
structures between $Q=\frac{6}{11}$ and $\frac{5}{9}$ are so close
in free energy below 3 K that the model needs to be modified in
order to differentiate clearly between the different $Q$ values in
this interval. This indicates that the peak in ${\mathcal J}({\bf
q})$ is possibly even more pronounced than assumed in the present
calculations. A strong peak in the RKKY-interaction may be
produced by the nesting between different areas on the Fermi
surface discussed by Dugdale {\it et al.},\cite{dugdale} a nesting
which is probably also responsible for the superconducting
properties of these compounds.

It is important to realize that the ferromagnetic transition in
ErNi$_2$B$_2$C is not due to a ferromagnetic interaction, which
one is actually strongly negative. The ferromagnetic component is
a byproduct of the low-temperature commensurable structure. Even a
large change of ${\mathcal J}({\bf 0})$ only have a slight
influence on the transition, as the exchange-energy gain of this
phase relatively to the pure antiferromagnetic phase is determined
by the combined contribution of all the even harmonics. This means
that the influence of the superconducting electrons on this
transition, as for instance through the Anderson--Suhl
mechanism,\cite{norgaard,anderson} is weak. When the applied field
in Fig.\ 2 is smaller than 1--2 kOe the response of the system is
diamagnetic. The width of the fundamental harmonic in Fig.\ 3
indicates a correlation length along [1,0,0] which is at least 80
$a$, about twice the superconducting coherence length $\xi$. The
absence of a spontaneous vortex phase, i.e.\ the diamagnetic
response of the system at low fields, therefore suggests that the
magnetic correlation length perpendicular to ${\bf Q}$ in the
Meissner phase is much shorter than the parallel component. The
model is going to be examined more closely in a paper presenting a
neutron-diffraction determination of the magnetic structures in an
applied field.\cite{toft}

Valuable discussions with P. Hedeg\aa rd, K. N\o rgaard Toft, and
N. H. Andersen are gratefully acknowledged.

\end{document}